\documentclass[prd,preprintnumbers,amssymb,amsmath,letterpaper,nofootinbib,twocolumn,floatfix]{revtex4-2}

\usepackage[dvipsnames]{xcolor}
\usepackage[colorlinks,allcolors=RoyalBlue]{hyperref}
\usepackage{graphicx}
\usepackage{bm}
\usepackage{microtype}

\newcommand{\Lavg}{{\langle L_\gamma \rangle}}

\makeatletter
\def\bibsection{%
   \par
   \begingroup
    \baselineskip26\p@
    \bib@device{\hsize}{72\p@}%
   \endgroup
   \nobreak\@nobreaktrue
   \addvspace{19\p@}%
  }%
\makeatother

\begin{document}

\preprint{\tt FERMILAB-PUB-23-583-T}

\title{A New Determination of the Millisecond Pulsar Gamma-Ray Luminosity Function and Implications for the Galactic Center Gamma-Ray Excess}

\author{Ian Holst$^{1,2}$}
\thanks{holst@uchicago.edu, https://orcid.org/0000-0003-4256-3680}

\author{Dan Hooper$^{1,2,3}$}
\thanks{dhooper@fnal.gov, http://orcid.org/0000-0001-8837-4127}

\affiliation{$^1$University of Chicago, Department of Astronomy and Astrophysics, Chicago, Illinois 60637, USA}
\affiliation{$^2$University of Chicago, Kavli Institute for Cosmological Physics, Chicago, Illinois 60637, USA}
\affiliation{$^3$Fermi National Accelerator Laboratory, Theoretical Astrophysics Group, Batavia, Illinois 60510, USA}

\date{\today}

\begin{abstract}
It has been suggested that the Galactic Center Gamma-Ray Excess (GCE) could be produced by a large number of centrally-located millisecond pulsars. The fact that no such pulsar population has been detected implies that these sources must be very faint and very numerous. In this study, we use the contents of Fermi's recently released Third Pulsar Catalog (3PC) to measure the luminosity function of the millisecond pulsars in the Milky Way's Disk. We find that this source population exhibits a luminosity function with a mean gamma-ray luminosity of $\langle L_\gamma \rangle \sim 6 \times 10^{32} \, \mathrm{erg/s}$ (integrated above 0.1 GeV). If the GCE were generated by millisecond pulsars with the same luminosity function, we find that $\sim 20$ such sources from the Inner Galaxy population should have already been detected by Fermi and included in the 3PC. Given the lack of such observed sources, we exclude the hypothesis that the GCE is generated by pulsars with the same luminosity function as those in the Galactic Disk with a significance of $3.4\sigma$. We conclude that either less than 39\% of the GCE is generated by pulsars, or that the millisecond pulsars in the Inner Galaxy are at least 5 times less luminous on average than those found in the Galactic Disk.
\end{abstract}

\maketitle

\section{Introduction}

Data collected by the Fermi Gamma-Ray Space Telescope has been used to identify a bright and statistically significant excess of GeV-scale gamma rays from the direction of the Inner Galaxy, when compared to all existing models of astrophysical diffuse emission and known point sources~\cite{Cholis:2021rpp,DiMauro:2021raz} (for early work, see Refs.~\cite{Goodenough:2009gk,Hooper:2010mq,Hooper:2011ti,Abazajian:2012pn,Hooper:2013rwa,Gordon:2013vta,Daylan:2014rsa,Calore:2014xka,Fermi-LAT:2015sau}). The spectrum, angular distribution, and overall normalization of this Galactic Center Gamma-Ray Excess (GCE) are each consistent with those predicted from annihilating dark matter particles.

The leading astrophysical interpretation of the GCE is that it is generated by a large population of unresolved millisecond pulsars (MSPs), distributed in a highly concentrated and approximately spherical or boxy bulge pattern throughout the Inner Galaxy~\cite{Hooper:2010mq,Hooper:2011ti,Abazajian:2012pn,Gordon:2013vta,Cholis:2014lta,Yuan:2014rca,Petrovic:2014xra,Brandt:2015ula}. The primary motivation for this possibility is that the spectral shape of the GCE is similar to that observed from typical MSPs. 

MSPs are old neutron stars that were spun-up to rapid rates of rotation through the transfer of angular momentum from a stellar companion~\cite{1982Natur.300..728A,1994ARA&A..32..591P,Lorimer:2001vd,Lorimer:2008se,2010ApJ...715..335K}. In contrast to young pulsars, MSPs have weaker magnetic fields and therefore lose rotational kinetic energy more slowly, often remaining luminous for billions of years. The long lifetimes of MSPs have lead some to speculate that large numbers of such objects could exist in or around the Galactic Center. If a large number of pulsars were present in the Inner Galaxy, then depending on their luminosity, it might be expected that some of these objects would have already been detected by Fermi as individual gamma-ray point sources (or as individual radio sources). A quantitative assessment of this problem requires a measurement of the MSP luminosity function. With this in mind, several groups have studied known MSP populations in an effort to measure or constrain their gamma-ray luminosity functions~\cite{Cholis:2014noa,Hooper:2015jlu,Hooper:2016rap,Ploeg_2017,Bartels:2018xom,Ploeg:2020jeh,Dinsmore:2021nip}. Under the assumption that the Inner Galaxy's MSP population has a similar luminosity function to that of the MSPs observed elsewhere, one can use the measured luminosity function of those sources to place constraints on the fraction of the GCE that could originate from unresolved MSPs.

\begin{figure*}[t]
\includegraphics[width=\linewidth]{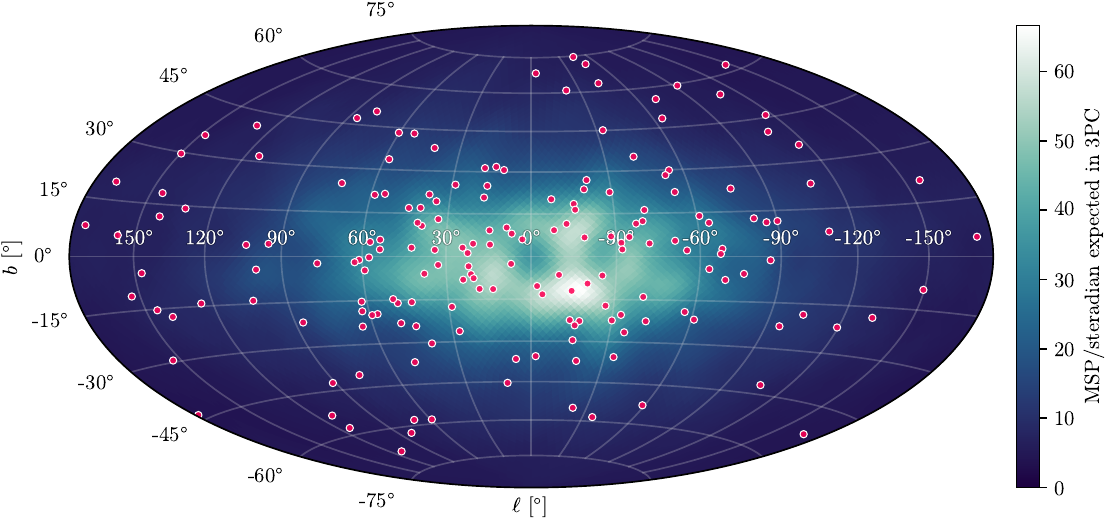} 
\caption{The red circles represent millisecond pulsars contained within Fermi's Third Pulsar Catalog (3PC), excluding those within globular clusters~\cite{Fermi-LAT:2023zzt}. This is overlaid on the angular MSP distribution that we predict should be observed by Fermi for our best-fit model parameters, summed over all flux bins.}
\label{fig:map}
\end{figure*}

The Fermi Collaboration recently published its Third Catalog of Gamma-Ray Pulsars (3PC)~\cite{Fermi-LAT:2023zzt}. This catalog includes 138 confirmed (TS $>$ 25) gamma-ray pulsars with millisecond-scale periods ($P < 30 \, {\rm ms}$), of which 4 are located in a globular cluster (the remaining 134 are distributed throughout the Galactic Disk, or the ``field'' of the Milky Way). In addition, the 3PC also contains 45 other gamma-ray sources which are co-located with a known MSP (see Tables~5 and~6 of Ref.~\cite{Fermi-LAT:2023zzt}); despite the lack of detected gamma-ray pulsations from these sources, they are likely to be MSPs. In this paper, we use the coordinates and gamma-ray fluxes of these 179 field MSPs (see Fig.~\ref{fig:map}) to measure the gamma-ray luminosity function of this source population. We then use this information to draw conclusions regarding the origin of the GCE. In particular, we find that if the GCE is generated by MSPs, the pulsars responsible for that signal must be significantly less luminous than those found within the Galactic Disk.

\bigskip

\section{Constraining The Luminosity Function of the Milky Way's Millisecond Pulsar Population} \label{sec:modeling}

We begin by constructing a model for the spatial distribution and luminosity function of the Milky Way's MSP population. For the spatial distribution, we adopt the following parameterization~\cite{2010JCAP...01..005F}:
\begin{equation} \label{eq:spatial_profile}
n_\mathrm{MSP} \propto e^{-R^2/2 \sigma_R^2} \, e^{-|z|/z_0},
\end{equation}
where $R$ and $z$ describe the location of a pulsar in cylindrical galactocentric coordinates, and $\sigma_R$ and $z_0$ are taken to be free parameters. In these coordinates, the Solar System is located at $R=8.122 \, {\rm kpc}$~\cite{GRAVITY:2018ofz} and $z=0.0208 \, {\rm kpc}$~\cite{2019MNRAS.482.1417B}. 

We parameterize the gamma-ray luminosity function of MSPs using a log-normal distribution,
\begin{equation}
\frac{dN}{dL_\gamma} \propto \frac{1}{L_\gamma} \exp\bigg(-
\frac{(\ln L_\gamma -\ln L_0)^2}{2 \sigma_L^2}\bigg),
\end{equation}
where we take $L_0$, $\sigma_L$, and the normalization $N_{\rm disk}$ (the total number of MSPs in the Galactic Disk) to be free parameters. Note that we will often present our results in terms of the mean gamma-ray luminosity, given by
\begin{equation}
\Lavg = L_0 \, \exp{\left( \frac{\sigma_L^2}{2} \right)}.
\end{equation}
We define a pulsar's gamma-ray luminosity $L_\gamma$ as the luminosity integrated above 0.1 GeV.

For a given set of model parameters, we can calculate the number of MSPs that should be included in the Fermi 3PC catalog in any direction on the sky:
\begin{equation} \label{eq:N3PC}
\begin{split}
N_{\rm 3PC} &= \int dL_\gamma \int dV \, n_\mathrm{MSP} \, \frac{dN}{dL_\gamma} \, P \\
&= \Delta \Omega \int dL_\gamma \int_{\rm los} dD \, D^2  \, n_\mathrm{MSP} \, \frac{dN}{dL_\gamma} \, P.
\end{split}
\end{equation}
The volume integral is performed over a cone of solid angle $\Delta \Omega$, and in the second line this is related to an integral over the line-of-sight distance $D$, with $dV= \Delta \Omega \, D^2 dD$.

\begin{figure*}[t]
\includegraphics[width=\linewidth]{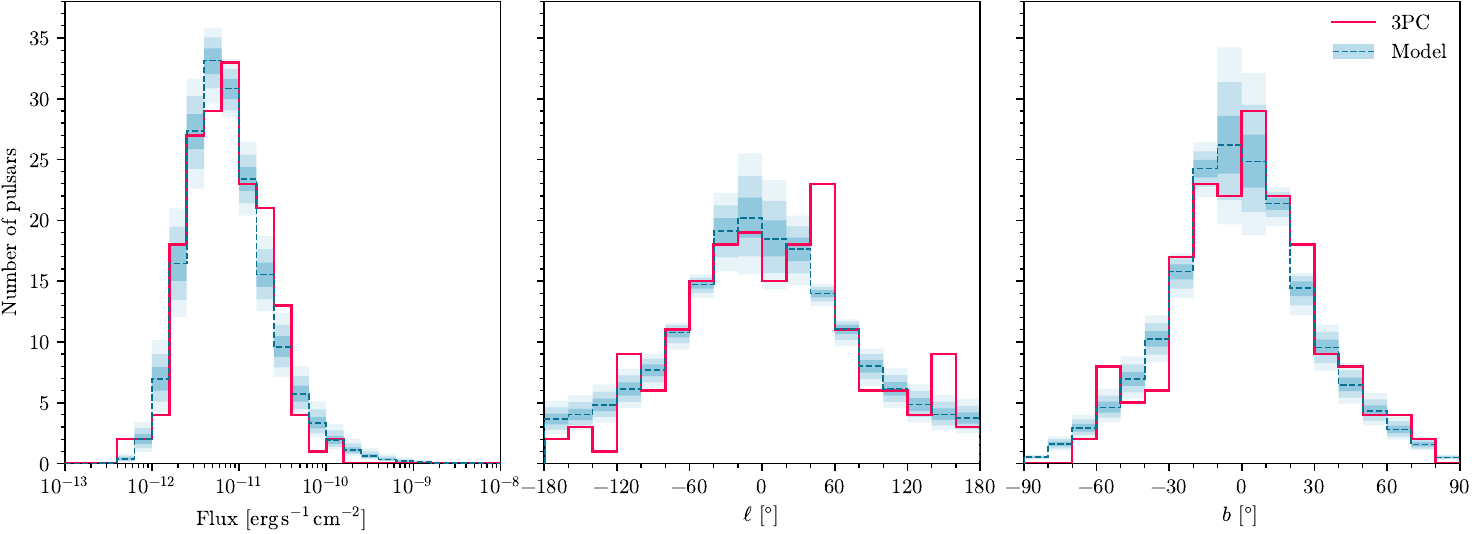}
\caption{The number of millisecond pulsars (MSPs) found in the Fermi Third Pulsar Catalog, in flux, longitude, and latitude bins (solid red lines). This is compared to the distributions that are predicted by our model with parameter values which lie within $2\Delta \ln \mathcal{L} =$ 1, 4 or 9 of the best-fit point (blue dashed line and surrounding bands).}
\label{fig:bins}
\end{figure*}

The factor $P$ is the probability that a pulsar with flux $F_{\gamma}=L_{\gamma}/4 \pi D^2$ in a given direction will be detected by Fermi and included in the 3PC catalog, taking into account Fermi's detection threshold for point sources. This threshold varies significantly with direction, being higher in directions with large backgrounds, such as along the Galactic Plane. To account for this, we use the sensitivity map as given in Fig.~25 of Ref.~\cite{Fermi-LAT:2023zzt}. As explained in that paper, this map can be rescaled by a numerical constant $\kappa$ to become a direction-dependent 50\% completeness threshold for inclusion in the 3PC catalog. To account for source-to-source variation, we take the detection threshold to be drawn from a log-normal distribution with a width of $\sigma_F$, and centered around the reported sensitivity. We fix these parameters to values which allow us to best-fit the observed pulsar flux distribution near Fermi's detection threshold, $\kappa=5.1$ and $\sigma_F=0.64$. Integrating over the log-normal distribution, we find:
\begin{equation}
    P(F_\gamma) = \frac{1}{2} \left[ 1 + \mathrm{erf} \left( \frac{ \ln F_{\gamma} - \ln (\kappa \, F_\mathrm{3PC}) }{\sqrt{2} \sigma_F} \right)  \right],
\end{equation}
where $F_{\mathrm{3PC}}$ is the 3PC sensitivity in a given direction.

For each set of model parameters, we compute the integral in Eq.~(\ref{eq:N3PC}) in each flux bin (using 5 flux bins per decade) and in each of the 12,288 angular pixels (a HEALPix map with \texttt{NSIDE}=32), convolving the result with a $5^{\circ}$ radius top-hat function to reduce the impact of small-scale variations in the sensitivity map. To downsample the original 3PC sensitivity map (HEALPix with \texttt{NSIDE}=256), for a given pulsar flux $F_\gamma$, we calculate the value of $P$ for each of the original 786,432 pixels. We then bin and average over this map to obtain a map with \texttt{NSIDE} = 32.

\begin{figure*}[t]
\includegraphics[width=0.49\linewidth]{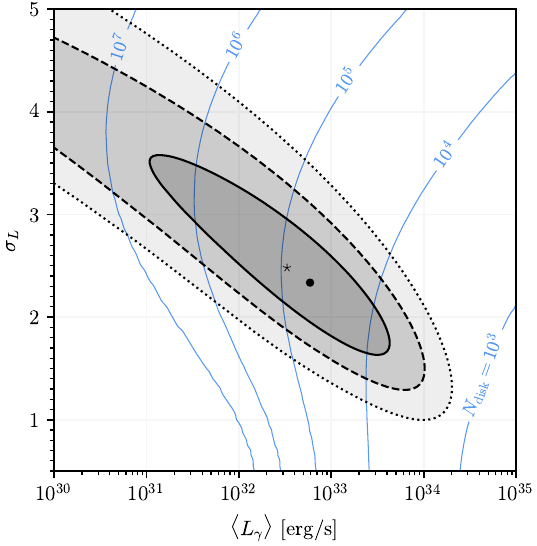}
\includegraphics[width=0.49\linewidth]{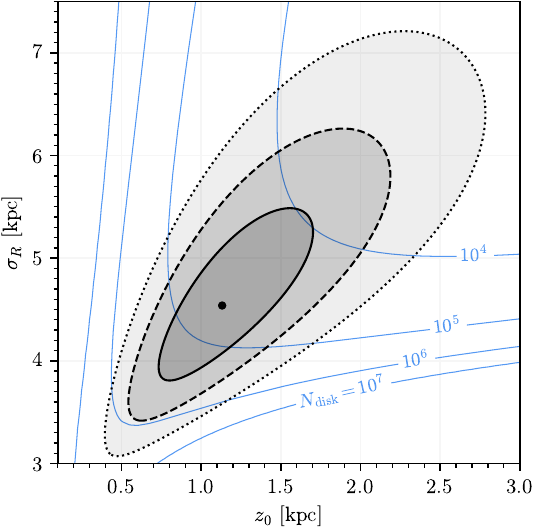}
\caption{Our determination of the parameters describing the gamma-ray luminosity function (left) and the spatial distribution (right) of the millisecond pulsar (MSP) population associated with the disk of the Milky Way (as opposed to those contained within globular clusters, or within our galaxy's bulge or central stellar cluster). The black dot denotes the best-fit point, while the solid, dashed, and dotted contours represent the 1$\sigma$, 2$\sigma$, and 3$\sigma$ confidence intervals, respectively. In each frame, we have marginalized over the unplotted parameters. The blue contours in each frame indicate the total number of disk MSPs, $N_{\rm disk}$. The asterisk in the left panel indicates the best-fit luminosity function for a Lorimer disk profile.}
\label{fig:param_constraints}
\end{figure*}

We then compare these predictions to the distribution of MSPs found within the 3PC catalog. The likelihood of some set of model parameters given this data is
\begin{equation}
\mathcal{L} =  \prod_i \frac{M_i^{D_i} e^{-M_i}}{D_i!},
\end{equation}
where $M_i$ and $D_i$ denote, respectively, the numbers of MSPs predicted by our model and observed by Fermi in the $i$th bin (of direction and flux). 

The best fit to the 3PC data is found for a luminosity function with $\Lavg \approx 5.9\times 10^{32}\, \mathrm{erg/s}$ and $\sigma_L \approx 2.3$ (corresponding to $L_0 \approx 3.9 \times 10^{31} \, \mathrm{erg/s}$). In Fig.~\ref{fig:map}, we show the locations of the 179 observed field MSPs contained within the 3PC overlaid on the distribution predicted for our best-fit model parameters, summed over all flux bins.

\begin{figure}
\includegraphics[width=\linewidth]{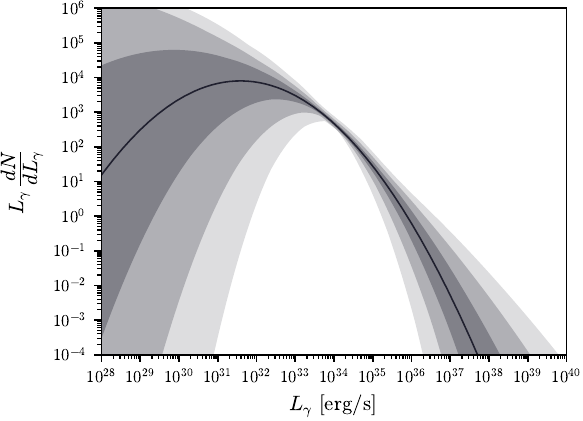} 
\caption{The range of millisecond pulsar (MSP) gamma-ray luminosity functions that are consistent (at the 1, 2, and 3$\sigma$ level) with the contents of Fermi's Third Pulsar Catalog (3PC).}
\label{fig:luminosity_function}
\end{figure}

There is, however, a significant range of parameter space around this best-fit point that is consistent with the 3PC data. In Fig.~\ref{fig:bins}, we show the distribution of MSPs in flux and angular bins, as found in the 3PC (solid red lines), compared to the distributions predicted for models with parameter values which yield fits which lie within $2\Delta \ln \mathcal{L} =$ 1, 4 or 9 of the best-fit parameter set (dashed line and shaded blue bands). In Fig.~\ref{fig:param_constraints}, we plot the regions of parameter space that are favored by our fit at the 1$\sigma$, 2$\sigma$, and $3\sigma$ confidence level (precise marginalized values are given in Table~\ref{tab:param_constraints}). In the left frame of this figure, we show our results in the $\Lavg$-$\sigma_L$ plane, marginalizing over the values of $\sigma_R$ and $z_0$, while in the right frame we show the results in the $\sigma_R$-$z_0$ plane, marginalizing over $\Lavg$ and $\sigma_L$ (in each case, the total number of disk MSPs, $N_{\rm disk}$, is fixed to yield a total of 179 observed MSPs). Note that for 2 degrees of freedom, these contours correspond to $-2\Delta \ln \mathcal{L} = 2.30$, 6.18, and 11.83 relative to the best-fit point. Also shown in this figure are contours representing different values of $N_{\rm disk}$. In Fig.~\ref{fig:luminosity_function}, we plot the MSP gamma-ray luminosity function over the range of parameters which yield $2\Delta \ln \mathcal{L}$ within 1, 2, or 9 of the best-fit point. Note that throughout this study we have restricted our luminosity function to parameter space with $\Lavg \ge 10^{30} \, \rm {erg/s}$.\footnote{We can estimate the number of neutron stars in the Milky Way by integrating the Salpeter mass function~\cite{1955ApJ...121..161S,Chabrier:2003ki}, $n(m) \propto m^{-2.35}$, between 8 and 25 $M_{\odot}$, finding that only 0.2\% of all stars have an initial mass in the range required to form a neutron star. This corresponds to an estimated $\sim 10^9$ neutron stars in the disk of the Milky Way. Given that only a small fraction of neutron stars will go on to become binary pulsars, we consider models with $N_{\rm disk} \gtrsim 10^7$ to be unrealistic.}

\begin{table*}[t]
\centering
\normalsize
\vspace{12pt}
\begin{tabular}{|cccccc|}
    \hline
    \textbf{Parameter} & \textbf{Best Fit} & $\bm{1 \sigma}$ \textbf{Range} & $\bm{2 \sigma}$ \textbf{Range} & $\bm{3 \sigma}$ \textbf{Range} & \textbf{Unit}\\
    \hline\hline
    $\Lavg$ & $5.9 \times 10^{32}$& $(0.67 - 23) \times 10^{32}$& $(0.01 - 67) \times 10^{32}$& $(0.01 - 150) \times 10^{32}$& erg/s\\
    $\sigma_L$ & $2.3$& $1.9 - 3.0$& $1.5 - 4.2$& $1.1 - 5.0$ & $-$ \\
    $z_0$ & $1.1$ & $0.87 - 1.5$& $0.65 - 1.9$& $0.47 - 2.5$& kpc\\
    $\sigma_R$ & $4.5$ & $4.1 - 5.1$& $3.7 - 5.8$& $3.2 - 6.7$& kpc\\
    $N_\mathrm{disk}$ & 47,000& 11,000 $-$ 470,000& 3,800 $-$ 39,000,000& 1,900 $-$ 56,000,000& $-$\\
    \hline
\end{tabular}
\caption{The parameter values favored by our analysis for the millisecond pulsar (MSP) luminosity function and spatial distribution.}
\label{tab:param_constraints}
\end{table*}

To test the model-dependence of our analysis, we also consider a Lorimer profile \cite{Lorimer:2006qs} to describe the spatial distribution of the disk pulsars. The Lorimer profile has three free parameters and can be defined as
\begin{equation}
    n_\mathrm{MSP} \propto R^B e^{-R/R_0} \, e^{-|z|/z_0}.
\end{equation}
We do not perform a full uncertainty analysis as with the Gaussian radial profile (Eq.~\ref{eq:spatial_profile}), but we do find the joint best-fit parameters for the Lorimer disk profile and log-normal luminosity function. The best-fit disk parameter values are $B \approx 3.7$, $R_0 \approx 1.1\,\mathrm{kpc}$, and $z_0 \approx 1.0\,\mathrm{kpc}$. The best-fit log-normal parameters are $\Lavg \approx 3.2\times 10^{32}\, \mathrm{erg/s}$ and $\sigma_L \approx 2.5$ (corresponding to $L_0 \approx 1.5 \times 10^{31} \, \mathrm{erg/s}$), also indicated on Fig.~\ref{fig:param_constraints}. Using a Lorimer disk profile improves the log-likelihood of the best-fit model over that of the best-fit using a Gaussian radial profile by a factor of $\Delta \ln \mathcal{L} = 1.2$. We consider that the addition of one additional parameter makes this not statistically significant.

\section{Comparisons With Previous Studies}

Several groups have attempted to measure the gamma-ray luminosity function of MSPs. In this section, we will discuss these previous results and compare them to the determination of the MSP luminosity function as presented in this study.

\begin{figure}[t]
\includegraphics[width=\linewidth]{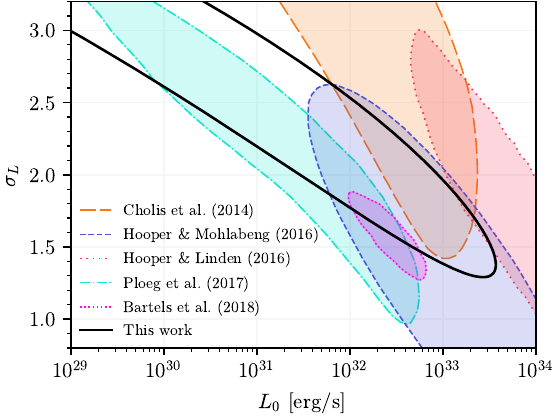}  
\caption{A comparison of our determination of the millisecond pulsar (MSP) gamma-ray luminosity function (solid black) to those presented in earlier studies~\cite{Cholis:2014noa,Hooper:2015jlu,Hooper:2016rap,Ploeg_2017,Bartels:2018xom}, each at the $2\sigma$ confidence level. The luminosity function found here is broadly consistent with these earlier results. Note that the study of Hooper \& Linden~\cite{Hooper:2016rap} was based on observations of millisecond pulsars in globular clusters, which could (in principle) feature a different luminosity function than pulsars in the Disk of the Milky Way.\label{fig:comparison_lognormal}}
\end{figure}

The analysis procedure followed in this work most closely resembles that carried out by Hooper~\& Mohlabeng in Ref.~\cite{Hooper:2015jlu}. Whereas this study utilized a dataset consisting of 179 MSPs in the field of the Milky Way, however, Ref.~\cite{Hooper:2015jlu} relied on a significantly smaller sample of 66 MSPs from the 3FGL catalog~\cite{Fermi-LAT:2015bhf}. As shown in Fig.~\ref{fig:comparison_lognormal}, our results are consistent with those presented in Ref.~\cite{Hooper:2015jlu}, in that our preferred regions of parameter space overlap substantially.

If we had reliable measurements of the distances to each of the pulsars in our sample, we could directly relate the measured fluxes of these sources to their intrinsic luminosities. Pulsar distance determinations, however, are often subject to significant and difficult to quantify systematic uncertainties. In particular, most known MSPs do not have measured values of their parallax. For the bulk of pulsars, distance estimates are instead derived from measurements of their dispersion measure ({\it ie}., the frequency-dependent time delay of a pulsar's radio pulses, which can be used to determine the the column density of free electrons along the pulsar's line-of-sight). When a pulsar's dispersion measure is combined with a model of the free electron distribution, this information can be used to produce a distance estimate. Such estimates can vary significantly, however, depending on the free electron distribution model that is adopted (for example, those of Refs.~\cite{2017ApJ...835...29Y,Cordes:2002wz}). Such distance estimates have sometimes been found to disagree with more reliable parallax determinations by factors of up to $\sim 3$ (see Ref.~\cite{Fermi-LAT:2023zzt} and references therein).

In light of these considerations, we have not used any pulsar distance measurements in our main analysis. In contrast, the earlier studies of Cholis {\it et al}.~\cite{Cholis:2014noa} and Bartels \textit{et al.}~\cite{Bartels:2018xom} each incorporated distance measurements into their determinations of the MSP luminosity function. By excluding distance information from our analysis, we are able to obtain a robust and reliable determination of the MSP luminosity function, albeit with reduced precision. In this sense, the approach taken here should be viewed as conservative.

In Fig.~\ref{fig:comparison_lognormal}, we compare our results to those of several previous studies. Our luminosity function determination is broadly consistent with the results of Cholis \textit{et al.}~\cite{Cholis:2014noa}\footnote{These curves are taken from \cite{Hooper:2015jlu} but are based on data from \cite{Cholis:2014noa}.}, Hooper~\& Mohlabeng~\cite{Hooper:2015jlu}, Ploeg \textit{et al.}~\cite{Ploeg_2017}, and Bartels \textit{et al.}~\cite{Bartels:2018xom}\footnote{A Lorimer spatial profile is used in the analysis of \cite{Bartels:2018xom}.}. Ref.~\cite{Ploeg:2020jeh} did not present their results in terms of a log-normal distribution, so they could not be included in our Fig.~\ref{fig:comparison_lognormal}, but broadly speaking, our results are statistically consistent with those of that study. We also show in Fig.~\ref{fig:comparison_lognormal} the range of luminosity function parameters that were favored in the study of Hooper \& Linden~\cite{Hooper:2016rap}, which was based on observations of the gamma-ray emission from globular clusters. The difference between our results and those of that study could, in principle, be due to differences between the luminosity functions of these two pulsars populations.

\section{Implications for the Galactic Center Gamma-Ray Excess}

Motivated by the goal of determining whether the GCE could be generated by a population of unresolved MSPs~\cite{Hooper:2010mq,Hooper:2011ti,Abazajian:2012pn,Gordon:2013vta,Yuan:2014rca,Petrovic:2014xra,Brandt:2015ula,Bartels:2015aea}, we adopt the range of luminosity functions identified in Sec.~\ref{sec:modeling}, and use this information to calculate the number of MSPs in the Inner Galaxy that should have already been detected by Fermi, assuming that pulsars with the same luminosity function as field MSPs are responsible for the observed gamma-ray excess. In this calculation, we adopt a spherically symmetric distribution of Inner Galaxy pulsars, with a number density that scales as $n \propto r^{-2.4}$, as favored by the observed morphology of the GCE~\cite{Cholis:2021rpp,Zhong:2024vyi}. We take this population to extend out to a distance of 3 kpc from the Galactic Center. We then normalize this population such that its total gamma-ray luminosity is equal to $L_{\gamma}^{\rm tot} = 2.1 \times 10^{37} \, \mathrm{erg/s}$, chosen to match the observed intensity of the GCE from within the innermost 3 kpc~\cite{Cholis:2021rpp}\footnote{In our calculations, we adopt the central value for the observed luminosity of the GCE. If the uncertainty on this quantity as reported in Ref.~\cite{Cholis:2021rpp} is taken into account, $L^{\rm tot}_{\gamma} = (2.1 \pm 0.3) \times 10^{37} \, {\rm erg/s}$, the uncertainty on our predicted values for $N_{\rm GCE, 3PC}$ will increase by a corresponding degree.}.

\begin{figure}
\includegraphics[width=\linewidth]{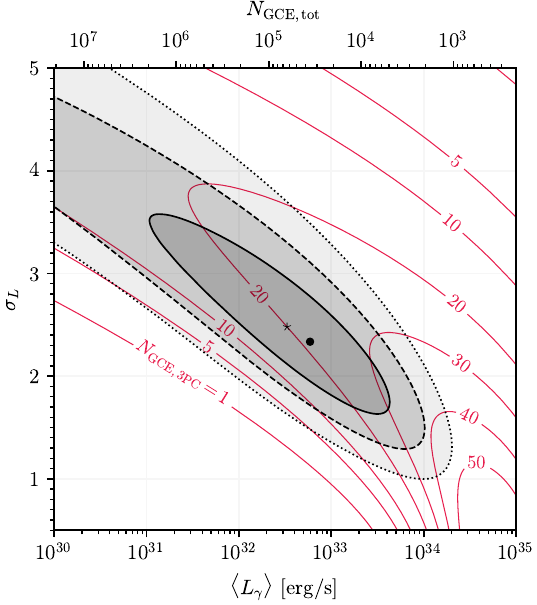}  
\caption{As in Fig.~\ref{fig:param_constraints}, but with contours showing the number of Inner Galaxy millisecond pulsars (MSPs) that should appear in the Fermi 3PC, assuming that the Galactic Center Gamma-Ray Excess (GCE) is generated by MSPs. The top axis indicates the conversion to the total number of millisecond pulsars that would be required to generate the total luminosity of the GCE. As there are only 3 Inner Galaxy pulsars candidates in the 3PC, MSPs can generate the excess gamma-ray emission only if that source population consists of members that are significantly less luminous than those present in the Galactic Plane.}
\label{fig:GCE_pulsar_constraints}
\end{figure}

\begin{table*}[t]
\centering
\normalsize
\begin{tabular}{|ccccc|}
    \hline
    \textbf{Parameter} & \textbf{Best Fit} & $\bm{1 \sigma}$ \textbf{Range} & $\bm{2 \sigma}$ \textbf{Range} & $\bm{3 \sigma}$ \textbf{Range} \\
    \hline\hline
    $N_\mathrm{GCE,\,tot}$ & 36,000& 9,300 $-$ 320,000& 3,200 $-$ 21,000,000& 1,400 $-$ 21,000,000\\
    $N_\mathrm{GCE,\,3PC}$ & $22$& $16-28$& $10-35$& $5.6-42$\\
    \hline
    $f_\mathrm{GCE}$ & $-$ & $<0.21$ & $<0.39$ & $<0.74$\\
    $\Lavg_\mathrm{disk} \, / \, \Lavg_\mathrm{IG}$ & $-$ & $>14$ & $>4.6$ & $>1.6$\\
    \hline
\end{tabular}
\caption{The total number of millisecond pulsars (MSPs) that would be required to produce the Galactic Center Gamma-Ray Excess, $N_\mathrm{GCE,\,tot}$, and the number of those pulsars that should be contained in Fermi's Third Pulsar Catalog (3PC), $N_\mathrm{GCE,\,tot}$, for the luminosity function parameters shown in Fig.~\ref{fig:param_constraints}. As there are only 3 Inner Galaxy pulsars candidates in the 3PC, MSPs can generate the excess gamma-ray emission only if that source population consists of members that are significantly less luminous than those present in the Galactic Plane. The last two rows show constraints on the maximum fraction of the gamma-ray excess that could be generated by MSPs, $f_{\rm GCE}$, assuming a common luminosity function, and the minimum ratio of the average luminosity of disk MSPs to Inner Galaxy MSPs, $\Lavg_\mathrm{disk} \, / \, \Lavg_\mathrm{IG}$.}
\label{tab:param_constraints_GCE}
\end{table*}

For this hypothesized pulsar population, we follow the approach described in Sec.~\ref{sec:modeling} to calculate how many MSPs from the Inner Galaxy population should be contained within Fermi's 3PC catalog. In Fig.~\ref{fig:GCE_pulsar_constraints}, we plot contours of the number of MSPs in the Inner Galaxy population that should be included in the 3PC, as a function of $\Lavg$ and $\sigma_L$. For our best-fit luminosity function parameters ($\Lavg = 5.9 \times 10^{32}\,\mathrm{erg/s}$, $\sigma_L=2.3$), we find that $N_\mathrm{GCE,\,tot} = 36,000$ MSPs would be required to generate the total measured intensity of the GCE. On average, $N_\mathrm{GCE,\,3PC} = 22$ of these MSPs should be bright enough to have been included within the 3PC catalog. Moving beyond the best-fit parameters, we find that for parameters within the 1$\sigma$ (2$\sigma$) range of our analysis, the 3PC catalog should contain between $N_\mathrm{GCE,\,3PC} = 16-28 \,\,(10-35)$ Inner Galaxy MSPs (see Table~\ref{tab:param_constraints_GCE}).

To assess the model-dependence of these results, we also considered a Coleman Bulge GCE profile \cite{Coleman:2019kax,Zhong:2024vyi} with a nuclear star cluster (NSC) \cite{2002A&A...384..112L,Bartels:2017vsx,Zhong:2024vyi}. We normalize the Coleman Bulge profile to a total gamma-ray luminosity of $2.3 \times 10^{37} \, \mathrm{erg/s}$, and the NSC profile to $3.9 \times 10^{36} \, \mathrm{erg/s}$ \cite{Macias:2019omb}. The sum of the two is 25\% brighter than the GCE luminosity that we use to normalize the spherical profile. Since the Coleman Bulge and NSC profiles are given projected on the sky, we assumed that all pulsars are at a distance of 8.122 kpc, unlike the spherical GCE profile which we integrate over the full 3D shape. In addition to our main results, which assumed a Gaussian disk profile and spherical GCE profile, we considered the following combinations:
\begin{itemize}
    \item Gaussian Disk + Coleman Bulge/NSC: 18 MSPs are expected to be included in the 3PC for the best-fit LF. The 1, 2, and 3$\sigma$ ranges of the LF parameters allow for $12 - 23$, $7.6 - 29$, and $3.8 - 35$ MSPs in the 3PC, respectively. A total of 45,000 pulsars would be present in the Inner Galaxy population.
    
    \item Lorimer Disk + Spherical GCE: 20 MSPs are expected to be included in the 3PC for the best-fit LF, with 63,000 total required to generate the GCE.
    
    \item Lorimer Disk + Coleman Bulge/NSC: 16 MSPs should be included in the 3PC for the best-fit LF. A total of 80,000 pulsars would be present in the Inner Galaxy population.
\end{itemize}

These predictions can be compared to the number of MSP candidates that Fermi has actually detected from the direction of the Inner Galaxy. In the 3PC catalog, there are several such candidates with appropriate values of $b$ and $l$, although most of these have distance measurements which place them much too far from the Galactic Center to be part of any source population that could be responsible for the GCE. There are, however, three known MSPs that could, in principle, be part of a central source population. These are PSR J1649-3012, PSR J1747-4036, and PSR J1833-3840, whose dispersion measures place them at distances of $d < 15.8 \, {\rm kpc}$, $d=7.14 \pm  2.85 \, {\rm kpc}$, and $d = 4.64 \pm 1.85 \, {\rm kpc}$, respectively~\cite{Fermi-LAT:2023zzt}. Combining this with the galactic coordinates of these pulsars, we find that they could each potentially be located within 3 kpc of the Galactic Center and be part of an Inner Galaxy population (they could also, of course, be part of the disk population).

One might further wonder whether there could be gamma-ray pulsars in the Inner Galaxy that have been detected as point sources by Fermi, but that have not yet had any pulsations detected, either in the gamma-ray or radio bands. With this in mind, we have reviewed the contents of the Fermi 4FGL catalog, finding no unassociated sources in the direction of the Inner Galaxy that have a pulsar-like spectral shape and that do not exhibit variability.\footnote{More precisely, for us to consider an unassociated Fermi source to be a pulsar candidate, we require its variability index to be less than 18.48, and its spectrum to be characterized by an exponentially cutoff power-law (about 90\% of 3PC pulsars are)} After applying these requirements, the only known gamma-ray sources that could be part of an Inner Galaxy MSP population are those mentioned in the previous paragraph (PSR J1649-3012, PSR J1747-4036, and PSR J1833-3840).

Combining our constraints on the MSP luminosity function with the fact that only three or fewer Inner Galaxy MSPs have been observed, we can exclude the possibility that the entire GCE is generated by a pulsar population with the same luminosity function as the field population at a level of 3.4$\sigma$ (using a Coleman Bulge with NSC profile slightly reduces the tension to 3.0$\sigma$). Alternatively, we can consider hypotheses in which only a fraction of the GCE is generated by MSPs. We constrain that fraction to $f_{\rm GCE} < 0.39$ (at the 95\% confidence level; see Table~\ref{tab:param_constraints_GCE}).

In light of the lack of Inner Galaxy MSP candidates contained in the 3PC catalog, pulsar interpretations of the GCE are in significant tension with the luminosity function that we have derived here. Despite this apparent tension, the results of this study do not strictly rule out the possibility that MSPs could be responsible for the GCE. In particular, it remains possible that the MSPs in the Inner Galaxy could be systematically fainter than those found in the Galactic Disk, or in globular clusters. If the average gamma-ray luminosity of MSPs in the Inner Galaxy were $\gtrsim 5$ times lower than that of the MSPs in the Galactic Disk, this tension would be relieved (at the 95\% confidence level; see Table~\ref{tab:param_constraints_GCE}). Such a scenario would require $N_{\rm GCE, tot} \gtrsim 2 \times 10^5$ MSPs to produce the observed intensity of the GCE.

The most straightforward way to explain how a population of MSPs could be systematically lower in luminosity would be for it to be older (see, for example, Ref.~\cite{Ploeg:2020jeh}). Pulsars lose rotational kinetic energy through magnetic dipole breaking, causing their spindown power to evolve as follows:
\begin{equation}
\dot{E} \propto \bigg(1+\frac{t}{\tau}\bigg)^2,
\end{equation}
where the spindown timescale is related to the pulsar's period and magnetic field: 
\begin{equation}
\tau \equiv \frac{E}{\dot{E}}= \frac{P}{2\dot{P}} \simeq 4.4 \, {\rm Gyr} \times \bigg(\frac{P}{5 \, {\rm ms}}\bigg)^2 \, \bigg(\frac{3\times 10^8 \, {\rm G}}{B}\bigg)^2.
\end{equation}
Thus, over the course of several Gyr, the average luminosity of a MSP could potentially fall by a significant factor. Alternatively, it has been suggested that the Inner Galaxy population might include pulsars that were produced through accretion-induced collapse, potentially leading to a population of low-luminosity gamma-ray sources~\cite{Ploeg_2017,Gautam:2021wqn}.

In previous studies, machine learning techniques have been applied to the Fermi data in an effort to search for evidence of unresolved point sources that might contribute to the GCE~\cite{List:2021aer, List:2020mzd,Mishra-Sharma:2021oxe}. The results of these analyses can also be used to place upper limits on the number of gamma-ray bright MSPs that are present in the Inner Galaxy. For example, the analysis of Ref.~\cite{List:2021aer} finds that a significant fraction of the GCE could arise from a population of $\sim 3 \times 10^4$ point sources~\cite{List:2021aer} (corresponding to $\Lavg \sim 7 \times 10^{32} \, \mathrm{erg/s}$). For this value of $\Lavg$, we would expect significantly more than three MSP candidates in the 3PC, unless $\sigma_L \lesssim 1.3$, which is well outside of the range favored by our analysis. 

Our results and the results of Refs.~\cite{List:2021aer,List:2020mzd,Mishra-Sharma:2021oxe} would each appear to be consistent with a scenario in which a relatively small fraction of the GCE ($f_{\rm GCE} \lesssim 0.39$) is generated by MSPs with our best-fit luminosity function. This scenario is also independently supported by the observed numbers of low-mass X-ray binaries in the Inner Galaxy and in globular clusters, which yields an estimate of $f_{\rm GCE} \sim 0.04-0.23$~\cite{Haggard:2017lyq}.

\section{Summary and Conclusions}

Whether or not millisecond pulsars (MSPs) could potentially generate the observed Galactic Center Gamma-Ray Excess (GCE) depends critically on the luminosity function, spatial distribution, and abundance of that source population. In this study, we have revisited the MSP gamma-ray luminosity function, making use of Fermi's recently released Third Pulsar Catalog (3PC). Comparing the contents of this catalog to the predicted distribution of MSPs, we have obtained a robust determination of the gamma-ray luminosity function for the population of MSPs present in the Milky Way's Disk.

The results presented here have significant implications for the origin of the GCE. In particular, if the Inner Galaxy's MSP population has a luminosity function that is similar to that derived in this study, we predict that $\sim 20$ of those pulsars should have already been detected by Fermi and be contained within the 3PC catalog. In actuality, there exist only three such pulsar candidates. From these arguments, we exclude the hypothesis that the GCE is generated by pulsars with the same luminosity function as those in the Galactic Disk with a significance of $3.4\sigma$. We also show that these results do not significantly depend on the choice of profile for either the Galactic Disk pulsars or the Inner Galaxy pulsars. These results can be used to conclude (at the 95\% confidence level) that either less than 39\% of the GCE is generated by pulsars, or that the Inner Galaxy MSPs are, on average, at least 5 times less luminous than those found in the Galactic Disk.

\vspace{16pt}

\textbf{Acknowledgments.} We thank Tim Linden, David Smith, Toby Burnett, Ilias Cholis, and Oscar Macias for helpful discussions. IH is supported by generous contributions from Philip Rice. DH is supported by the Fermi Research Alliance, LLC under Contract No.~DE-AC02-07CH11359 with the U.S. Department of Energy, Office of Science, Office of High Energy Physics.

\bibliographystyle{utphys}
\bibliography{msp2023}

\providecommand{\href}[2]{#2}\begingroup\raggedright\begin{thebibliography}{10}

\bibitem{Cholis:2021rpp}
I.~Cholis, Y.-M. Zhong, S.~D. McDermott, and J.~P. Surdutovich, ``{Return of the templates: Revisiting the Galactic Center excess with multimessenger observations},'' \href{https://dx.doi.org/10.1103/PhysRevD.105.103023}{{\em Phys. Rev. D} {\bfseries 105} no.~10, (2022) 103023}, \href{https://arxiv.org/abs/2112.09706}{{\ttfamily arXiv:2112.09706 [astro-ph.HE]}}.

\bibitem{DiMauro:2021raz}
M.~Di~Mauro, ``{Characteristics of the Galactic Center excess measured with 11 years of $Fermi$-LAT data},'' \href{https://dx.doi.org/10.1103/PhysRevD.103.063029}{{\em Phys. Rev. D} {\bfseries 103} no.~6, (2021) 063029}, \href{https://arxiv.org/abs/2101.04694}{{\ttfamily arXiv:2101.04694 [astro-ph.HE]}}.

\bibitem{Goodenough:2009gk}
L.~Goodenough and D.~Hooper, ``{Possible Evidence For Dark Matter Annihilation In The Inner Milky Way From The Fermi Gamma Ray Space Telescope},'' \href{https://arxiv.org/abs/0910.2998}{{\ttfamily arXiv:0910.2998 [hep-ph]}}.

\bibitem{Hooper:2010mq}
D.~Hooper and L.~Goodenough, ``{Dark Matter Annihilation in The Galactic Center As Seen by the Fermi Gamma Ray Space Telescope},'' \href{https://dx.doi.org/10.1016/j.physletb.2011.02.029}{{\em Phys. Lett. B} {\bfseries 697} (2011) 412--428}, \href{https://arxiv.org/abs/1010.2752}{{\ttfamily arXiv:1010.2752 [hep-ph]}}.

\bibitem{Hooper:2011ti}
D.~Hooper and T.~Linden, ``{On The Origin Of The Gamma Rays From The Galactic Center},'' \href{https://dx.doi.org/10.1103/PhysRevD.84.123005}{{\em Phys. Rev. D} {\bfseries 84} (2011) 123005}, \href{https://arxiv.org/abs/1110.0006}{{\ttfamily arXiv:1110.0006 [astro-ph.HE]}}.

\bibitem{Abazajian:2012pn}
K.~N. Abazajian and M.~Kaplinghat, ``{Detection of a Gamma-Ray Source in the Galactic Center Consistent with Extended Emission from Dark Matter Annihilation and Concentrated Astrophysical Emission},'' \href{https://dx.doi.org/10.1103/PhysRevD.86.083511}{{\em Phys. Rev. D} {\bfseries 86} (2012) 083511}, \href{https://arxiv.org/abs/1207.6047}{{\ttfamily arXiv:1207.6047 [astro-ph.HE]}}. [Erratum: Phys.Rev.D 87, 129902 (2013)].

\bibitem{Hooper:2013rwa}
D.~Hooper and T.~R. Slatyer, ``{Two Emission Mechanisms in the Fermi Bubbles: A Possible Signal of Annihilating Dark Matter},'' \href{https://dx.doi.org/10.1016/j.dark.2013.06.003}{{\em Phys. Dark Univ.} {\bfseries 2} (2013) 118--138}, \href{https://arxiv.org/abs/1302.6589}{{\ttfamily arXiv:1302.6589 [astro-ph.HE]}}.

\bibitem{Gordon:2013vta}
C.~Gordon and O.~Macias, ``{Dark Matter and Pulsar Model Constraints from Galactic Center Fermi-LAT Gamma Ray Observations},'' \href{https://dx.doi.org/10.1103/PhysRevD.88.083521}{{\em Phys. Rev. D} {\bfseries 88} no.~8, (2013) 083521}, \href{https://arxiv.org/abs/1306.5725}{{\ttfamily arXiv:1306.5725 [astro-ph.HE]}}. [Erratum: Phys.Rev.D 89, 049901 (2014)].

\bibitem{Daylan:2014rsa}
T.~Daylan, D.~P. Finkbeiner, D.~Hooper, T.~Linden, S.~K.~N. Portillo, N.~L. Rodd, and T.~R. Slatyer, ``{The characterization of the gamma-ray signal from the central Milky Way: A case for annihilating dark matter},'' \href{https://dx.doi.org/10.1016/j.dark.2015.12.005}{{\em Phys. Dark Univ.} {\bfseries 12} (2016) 1--23}, \href{https://arxiv.org/abs/1402.6703}{{\ttfamily arXiv:1402.6703 [astro-ph.HE]}}.

\bibitem{Calore:2014xka}
F.~Calore, I.~Cholis, and C.~Weniger, ``{Background Model Systematics for the Fermi GeV Excess},'' \href{https://dx.doi.org/10.1088/1475-7516/2015/03/038}{{\em JCAP} {\bfseries 03} (2015) 038}, \href{https://arxiv.org/abs/1409.0042}{{\ttfamily arXiv:1409.0042 [astro-ph.CO]}}.

\bibitem{Fermi-LAT:2015sau}
{\bfseries Fermi-LAT} Collaboration, M.~Ajello {\em et~al.}, ``{Fermi-LAT Observations of High-Energy $\gamma$-Ray Emission Toward the Galactic Center},'' \href{https://dx.doi.org/10.3847/0004-637X/819/1/44}{{\em Astrophys. J.} {\bfseries 819} no.~1, (2016) 44}, \href{https://arxiv.org/abs/1511.02938}{{\ttfamily arXiv:1511.02938 [astro-ph.HE]}}.

\bibitem{Cholis:2014lta}
I.~Cholis, D.~Hooper, and T.~Linden, ``{Challenges in Explaining the Galactic Center Gamma-Ray Excess with Millisecond Pulsars},'' \href{https://dx.doi.org/10.1088/1475-7516/2015/06/043}{{\em JCAP} {\bfseries 06} (2015) 043}, \href{https://arxiv.org/abs/1407.5625}{{\ttfamily arXiv:1407.5625 [astro-ph.HE]}}.

\bibitem{Yuan:2014rca}
Q.~Yuan and B.~Zhang, ``{Millisecond pulsar interpretation of the Galactic center gamma-ray excess},'' \href{https://dx.doi.org/10.1016/j.jheap.2014.06.001}{{\em JHEAp} {\bfseries 3-4} (2014) 1--8}, \href{https://arxiv.org/abs/1404.2318}{{\ttfamily arXiv:1404.2318 [astro-ph.HE]}}.

\bibitem{Petrovic:2014xra}
J.~Petrovi\'c, P.~D. Serpico, and G.~Zaharijas, ``{Millisecond pulsars and the Galactic Center gamma-ray excess: the importance of luminosity function and secondary emission},'' \href{https://dx.doi.org/10.1088/1475-7516/2015/02/023}{{\em JCAP} {\bfseries 02} (2015) 023}, \href{https://arxiv.org/abs/1411.2980}{{\ttfamily arXiv:1411.2980 [astro-ph.HE]}}.

\bibitem{Brandt:2015ula}
T.~D. Brandt and B.~Kocsis, ``{Disrupted Globular Clusters Can Explain the Galactic Center Gamma Ray Excess},'' \href{https://dx.doi.org/10.1088/0004-637X/812/1/15}{{\em Astrophys. J.} {\bfseries 812} no.~1, (2015) 15}, \href{https://arxiv.org/abs/1507.05616}{{\ttfamily arXiv:1507.05616 [astro-ph.HE]}}.

\bibitem{1982Natur.300..728A}
M.~A. {Alpar}, A.~F. {Cheng}, M.~A. {Ruderman}, and J.~{Shaham}, ``{A new class of radio pulsars},'' \href{https://dx.doi.org/10.1038/300728a0}{{\em Nature} {\bfseries 300} no.~5894, (Dec., 1982) 728--730}.

\bibitem{1994ARA&A..32..591P}
E.~S. {Phinney} and S.~R. {Kulkarni}, ``{Binary and Millisecond Pulsars},'' \href{https://dx.doi.org/10.1146/annurev.aa.32.090194.003111}{{\em Annu.~Rev.~Astron.~Astrophys.} {\bfseries 32} (Jan., 1994) 591--639}.

\bibitem{Lorimer:2001vd}
D.~R. Lorimer, ``{Binary and millisecond pulsars at the new millennium},'' \href{https://dx.doi.org/10.12942/lrr-2001-5}{{\em Living Rev. Rel.} {\bfseries 4} (2001) 5}, \href{https://arxiv.org/abs/astro-ph/0104388}{{\ttfamily arXiv:astro-ph/0104388}}.

\bibitem{Lorimer:2008se}
D.~R. Lorimer, ``{Binary and Millisecond Pulsars},'' \href{https://dx.doi.org/10.12942/lrr-2008-8}{{\em Living Rev. Rel.} {\bfseries 11} (2008) 8}, \href{https://arxiv.org/abs/0811.0762}{{\ttfamily arXiv:0811.0762 [astro-ph]}}.

\bibitem{2010ApJ...715..335K}
B.~{Kiziltan} and S.~E. {Thorsett}, ``{Millisecond Pulsar Ages: Implications of Binary Evolution and a Maximum Spin Limit},'' \href{https://dx.doi.org/10.1088/0004-637X/715/1/335}{{\em Astrophys. J.} {\bfseries 715} no.~1, (May, 2010) 335--341}, \href{https://arxiv.org/abs/0909.1562}{{\ttfamily arXiv:0909.1562 [astro-ph.GA]}}.

\bibitem{Cholis:2014noa}
I.~Cholis, D.~Hooper, and T.~Linden, ``{A New Determination of the Spectra and Luminosity Function of Gamma-Ray Millisecond Pulsars},'' \href{https://arxiv.org/abs/1407.5583}{{\ttfamily arXiv:1407.5583 [astro-ph.HE]}}.

\bibitem{Hooper:2015jlu}
D.~Hooper and G.~Mohlabeng, ``{The Gamma-Ray Luminosity Function of Millisecond Pulsars and Implications for the GeV Excess},'' \href{https://dx.doi.org/10.1088/1475-7516/2016/03/049}{{\em JCAP} {\bfseries 03} (2016) 049}, \href{https://arxiv.org/abs/1512.04966}{{\ttfamily arXiv:1512.04966 [astro-ph.HE]}}.

\bibitem{Hooper:2016rap}
D.~Hooper and T.~Linden, ``{The Gamma-Ray Pulsar Population of Globular Clusters: Implications for the GeV Excess},'' \href{https://dx.doi.org/10.1088/1475-7516/2016/08/018}{{\em JCAP} {\bfseries 08} (2016) 018}, \href{https://arxiv.org/abs/1606.09250}{{\ttfamily arXiv:1606.09250 [astro-ph.HE]}}.

\bibitem{Ploeg_2017}
H.~Ploeg, C.~Gordon, R.~Crocker, and O.~Macias, ``{Consistency Between the Luminosity Function of Resolved Millisecond Pulsars and the Galactic Center Excess},'' \href{https://dx.doi.org/10.1088/1475-7516/2017/08/015}{{\em JCAP} {\bfseries 08} (2017) 015}, \href{https://arxiv.org/abs/1705.00806}{{\ttfamily arXiv:1705.00806 [astro-ph.HE]}}.

\bibitem{Bartels:2018xom}
R.~T. Bartels, T.~D.~P. Edwards, and C.~Weniger, ``{Bayesian model comparison and analysis of the Galactic disc population of gamma-ray millisecond pulsars},'' \href{https://dx.doi.org/10.1093/mnras/sty2529}{{\em Mon. Not. Roy. Astron. Soc.} {\bfseries 481} no.~3, (2018) 3966--3987}, \href{https://arxiv.org/abs/1805.11097}{{\ttfamily arXiv:1805.11097 [astro-ph.HE]}}.

\bibitem{Ploeg:2020jeh}
H.~Ploeg, C.~Gordon, R.~Crocker, and O.~Macias, ``{Comparing the Galactic Bulge and Galactic Disk Millisecond Pulsars},'' \href{https://dx.doi.org/10.1088/1475-7516/2020/12/035}{{\em JCAP} {\bfseries 12} (2020) 035}, \href{https://arxiv.org/abs/2008.10821}{{\ttfamily arXiv:2008.10821 [astro-ph.HE]}}. [Erratum: JCAP 07, E01 (2021)].

\bibitem{Dinsmore:2021nip}
J.~T. Dinsmore and T.~R. Slatyer, ``{Luminosity functions consistent with a pulsar-dominated Galactic Center excess},'' \href{https://dx.doi.org/10.1088/1475-7516/2022/06/025}{{\em JCAP} {\bfseries 06} no.~06, (2022) 025}, \href{https://arxiv.org/abs/2112.09699}{{\ttfamily arXiv:2112.09699 [astro-ph.HE]}}.

\bibitem{Fermi-LAT:2023zzt}
{\bfseries Fermi-LAT} Collaboration, D.~A. Smith {\em et~al.}, ``{The Third Fermi Large Area Telescope Catalog of Gamma-Ray Pulsars},'' \href{https://dx.doi.org/10.3847/1538-4357/acee67}{{\em Astrophys. J.} {\bfseries 958} no.~2, (2023) 191}, \href{https://arxiv.org/abs/2307.11132}{{\ttfamily arXiv:2307.11132 [astro-ph.HE]}}.

\bibitem{2010JCAP...01..005F}
C.-A. {Faucher-Gigu{\`e}re} and A.~{Loeb}, ``{The pulsar contribution to the gamma-ray background},'' \href{https://dx.doi.org/10.1088/1475-7516/2010/01/005}{{\em JCAP} {\bfseries 2010} no.~1, (Jan., 2010) 005}, \href{https://arxiv.org/abs/0904.3102}{{\ttfamily arXiv:0904.3102 [astro-ph.HE]}}.

\bibitem{GRAVITY:2018ofz}
{\bfseries GRAVITY} Collaboration, R.~Abuter {\em et~al.}, ``{Detection of the gravitational redshift in the orbit of the star S2 near the Galactic centre massive black hole},'' \href{https://dx.doi.org/10.1051/0004-6361/201833718}{{\em Astron. Astrophys.} {\bfseries 615} (2018) L15}, \href{https://arxiv.org/abs/1807.09409}{{\ttfamily arXiv:1807.09409 [astro-ph.GA]}}.

\bibitem{2019MNRAS.482.1417B}
M.~{Bennett} and J.~{Bovy}, ``{Vertical waves in the solar neighbourhood in Gaia DR2},'' \href{https://dx.doi.org/10.1093/mnras/sty2813}{{\em Mon. Not. Roy. Astron. Soc.} {\bfseries 482} no.~1, (Jan., 2019) 1417--1425}, \href{https://arxiv.org/abs/1809.03507}{{\ttfamily arXiv:1809.03507 [astro-ph.GA]}}.

\bibitem{1955ApJ...121..161S}
E.~E. {Salpeter}, ``{The Luminosity Function and Stellar Evolution.},'' \href{https://dx.doi.org/10.1086/145971}{{\em \apj} {\bfseries 121} (Jan., 1955) 161}.

\bibitem{Chabrier:2003ki}
G.~Chabrier, ``{Galactic stellar and substellar initial mass function},'' \href{https://dx.doi.org/10.1086/376392}{{\em Publ. Astron. Soc. Pac.} {\bfseries 115} (2003) 763--796}, \href{https://arxiv.org/abs/astro-ph/0304382}{{\ttfamily arXiv:astro-ph/0304382}}.

\bibitem{Lorimer:2006qs}
D.~R. Lorimer {\em et~al.}, ``{The Parkes multibeam pulsar survey: VI. Discovery and timing of 142 pulsars and a Galactic population analysis},'' \href{https://dx.doi.org/10.1111/j.1365-2966.2006.10887.x}{{\em Mon. Not. Roy. Astron. Soc.} {\bfseries 372} (2006) 777--800}, \href{https://arxiv.org/abs/astro-ph/0607640}{{\ttfamily arXiv:astro-ph/0607640}}.

\bibitem{Fermi-LAT:2015bhf}
{\bfseries Fermi-LAT} Collaboration, F.~Acero {\em et~al.}, ``{Fermi Large Area Telescope Third Source Catalog},'' \href{https://dx.doi.org/10.1088/0067-0049/218/2/23}{{\em Astrophys. J. Suppl.} {\bfseries 218} no.~2, (2015) 23}, \href{https://arxiv.org/abs/1501.02003}{{\ttfamily arXiv:1501.02003 [astro-ph.HE]}}.

\bibitem{2017ApJ...835...29Y}
J.~M. {Yao}, R.~N. {Manchester}, and N.~{Wang}, ``{A New Electron-density Model for Estimation of Pulsar and FRB Distances},'' \href{https://dx.doi.org/10.3847/1538-4357/835/1/29}{{\em Astrophys. J.} {\bfseries 835} no.~1, (Jan., 2017) 29}, \href{https://arxiv.org/abs/1610.09448}{{\ttfamily arXiv:1610.09448 [astro-ph.GA]}}.

\bibitem{Cordes:2002wz}
J.~M. Cordes and T.~J.~W. Lazio, ``{NE2001. 1. A New model for the galactic distribution of free electrons and its fluctuations},'' \href{https://arxiv.org/abs/astro-ph/0207156}{{\ttfamily arXiv:astro-ph/0207156}}.

\bibitem{Bartels:2015aea}
R.~Bartels, S.~Krishnamurthy, and C.~Weniger, ``{Strong support for the millisecond pulsar origin of the Galactic center GeV excess},'' \href{https://dx.doi.org/10.1103/PhysRevLett.116.051102}{{\em Phys. Rev. Lett.} {\bfseries 116} no.~5, (2016) 051102}, \href{https://arxiv.org/abs/1506.05104}{{\ttfamily arXiv:1506.05104 [astro-ph.HE]}}.

\bibitem{Zhong:2024vyi}
Y.-M. Zhong and I.~Cholis, ``{Robustness of the Galactic Center excess morphology against masking},'' \href{https://dx.doi.org/10.1103/PhysRevD.109.123017}{{\em Phys. Rev. D} {\bfseries 109} no.~12, (2024) 123017}, \href{https://arxiv.org/abs/2401.02481}{{\ttfamily arXiv:2401.02481 [astro-ph.HE]}}.

\bibitem{Coleman:2019kax}
B.~Coleman, D.~Paterson, C.~Gordon, O.~Macias, and H.~Ploeg, ``{Maximum Entropy Estimation of the Galactic Bulge Morphology via the VVV Red Clump},'' \href{https://dx.doi.org/10.1093/mnras/staa1281}{{\em Mon. Not. Roy. Astron. Soc.} {\bfseries 495} no.~3, (2020) 3350--3372}, \href{https://arxiv.org/abs/1911.04714}{{\ttfamily arXiv:1911.04714 [astro-ph.GA]}}.

\bibitem{2002A&A...384..112L}
R.~{Launhardt}, R.~{Zylka}, and P.~G. {Mezger}, ``{The nuclear bulge of the Galaxy. III. Large-scale physical characteristics of stars and interstellar matter},'' \href{https://dx.doi.org/10.1051/0004-6361:20020017}{{\em Astron. Astrophys.} {\bfseries 384} (Mar., 2002) 112--139}, \href{https://arxiv.org/abs/astro-ph/0201294}{{\ttfamily arXiv:astro-ph/0201294 [astro-ph]}}.

\bibitem{Bartels:2017vsx}
R.~Bartels, E.~Storm, C.~Weniger, and F.~Calore, ``{The Fermi-LAT GeV excess as a tracer of stellar mass in the Galactic bulge},'' \href{https://dx.doi.org/10.1038/s41550-018-0531-z}{{\em Nature Astron.} {\bfseries 2} no.~10, (2018) 819--828}, \href{https://arxiv.org/abs/1711.04778}{{\ttfamily arXiv:1711.04778 [astro-ph.HE]}}.

\bibitem{Macias:2019omb}
O.~Macias, S.~Horiuchi, M.~Kaplinghat, C.~Gordon, R.~M. Crocker, and D.~M. Nataf, ``{Strong Evidence that the Galactic Bulge is Shining in Gamma Rays},'' \href{https://dx.doi.org/10.1088/1475-7516/2019/09/042}{{\em JCAP} {\bfseries 09} (2019) 042}, \href{https://arxiv.org/abs/1901.03822}{{\ttfamily arXiv:1901.03822 [astro-ph.HE]}}.

\bibitem{Gautam:2021wqn}
A.~Gautam, R.~M. Crocker, L.~Ferrario, A.~J. Ruiter, H.~Ploeg, C.~Gordon, and O.~Macias, ``{Millisecond pulsars from accretion-induced collapse as the origin of the Galactic Centre gamma-ray excess signal},'' \href{https://dx.doi.org/10.1038/s41550-022-01658-3}{{\em Nature Astron.} {\bfseries 6} no.~6, (2022) 703--707}, \href{https://arxiv.org/abs/2106.00222}{{\ttfamily arXiv:2106.00222 [astro-ph.HE]}}.

\bibitem{List:2021aer}
F.~List, N.~L. Rodd, and G.~F. Lewis, ``{Extracting the Galactic Center excess\textquoteright{} source-count distribution with neural nets},'' \href{https://dx.doi.org/10.1103/PhysRevD.104.123022}{{\em Phys. Rev. D} {\bfseries 104} no.~12, (2021) 123022}, \href{https://arxiv.org/abs/2107.09070}{{\ttfamily arXiv:2107.09070 [astro-ph.HE]}}.

\bibitem{List:2020mzd}
F.~List, N.~L. Rodd, G.~F. Lewis, and I.~Bhat, ``{The GCE in a New Light: Disentangling the $\gamma$-ray Sky with Bayesian Graph Convolutional Neural Networks},'' \href{https://dx.doi.org/10.1103/PhysRevLett.125.241102}{{\em Phys. Rev. Lett.} {\bfseries 125} (2020) 241102}, \href{https://arxiv.org/abs/2006.12504}{{\ttfamily arXiv:2006.12504 [astro-ph.HE]}}.

\bibitem{Mishra-Sharma:2021oxe}
S.~Mishra-Sharma and K.~Cranmer, ``{Neural simulation-based inference approach for characterizing the Galactic Center \ensuremath{\gamma}-ray excess},'' \href{https://dx.doi.org/10.1103/PhysRevD.105.063017}{{\em Phys. Rev. D} {\bfseries 105} no.~6, (2022) 063017}, \href{https://arxiv.org/abs/2110.06931}{{\ttfamily arXiv:2110.06931 [astro-ph.HE]}}.

\bibitem{Haggard:2017lyq}
D.~Haggard, C.~Heinke, D.~Hooper, and T.~Linden, ``{Low Mass X-Ray Binaries in the Inner Galaxy: Implications for Millisecond Pulsars and the GeV Excess},'' \href{https://dx.doi.org/10.1088/1475-7516/2017/05/056}{{\em JCAP} {\bfseries 05} (2017) 056}, \href{https://arxiv.org/abs/1701.02726}{{\ttfamily arXiv:1701.02726 [astro-ph.HE]}}.

\end{thebibliography}\endgroup

\end{document}